\documentclass[11pt]{article} 

\usepackage{amsmath}

\usepackage[adobe-utopia]{mathdesign}
\usepackage{graphicx}
\usepackage{caption}
\usepackage[inner=1.5in, top = 0.8in]{geometry}
\usepackage[hidelinks]{hyperref}
\usepackage{cleveref}

\usepackage{natbib}

\usepackage[dvipsnames, table]{xcolor}
\definecolor{maincol}{rgb}{0, 0.29, 0.33}
\definecolor{linkcol}{RGB}{0, 0, 0}
\definecolor{newforMW}{rgb}{0, 0, 1}

\usepackage{lineno}
\usepackage{paralist}
\newenvironment{paraenum}{\begin{inparaenum}[(\itshape i\upshape)]}{\end{inparaenum}}

\usepackage{helvet}
\usepackage{titlesec}
\titleformat{\section}
  {\normalfont\Large\bfseries\sffamily\color{maincol}}{\thesection.}{1em}{}
\titleformat{\subsection}
  {\normalfont\large\bfseries\sffamily}{\thesubsection}{1em}{}
\titleformat{\subsubsection}
  {\normalfont\normalsize\itshape\sffamily}{\thesubsubsection.}{1em}{}

\makeatletter
    \def\@maketitle{%
  \newpage
  \null
  \vskip 2em%
  \let \footnote \thanks
   \noindent {\Large \bfseries \sffamily \color{maincol} \@title \par}%
    \vskip 1.5em%
    \noindent {\normalsize
        \@author
	\par}%
    \vskip 0.5em%
   {\scriptsize \noindent \@date \par
   }
  \par
  \vskip 1.5em}
\makeatother

\newcommand{\quotestyle}[1]{{``\emph{#1}''}}

\newcommand{\SC}[0]{SC2024}
\newcommand{\WHSM}[0]{W2022}
\newcommand{\Covid}{\mbox{COVID-19}}
\newcommand{\sct}{\mbox{SARS-CoV-2}}
\newcommand{\ecl}{December case locations}
\newcommand{\worldpop}{{world\-pop.org}}

\title{Confirmation of the centrality of the Huanan market among early \\\Covid{} cases \\ {\large \sf {Reply to \citet{StoyanChiu2024}}}}

\author{F. D\'ebarre$^1$ \& M. Worobey$^2$}
\date{}
\begin{document}
\maketitle

{\noindent \small $^1$ 
Institut d’Écologie et des Sciences de l’Environnement (IEES-Paris, UMR 7618), CNRS, Sorbonne Université, UPEC, IRD, INRAE, Paris, France. ORCID: 0000-0003-2497-833X.\\ Contact: \mbox{florence.debarre@sorbonne-universite.fr} \\
$^2$ Department of Ecology and Evolutionary Biology, University of Arizona, Tucson, AZ, USA. \\ Contact: \mbox{worobey@arizona.edu}}

% Input numerical values
\def\hall{866}
\def\hlinked{844}
\def\hunlinked{952}
\def\nreps{1999}
\def\nrepsfirstinf{10^{6}}
\def\ncases{155}
\def\nWeiboXuJan{440}
\def\nWeiboXuFeb{233}
\def\nlinked{35}
\def\nunlinked{120}
\def\Hmat{\begin{pmatrix} 2.8\times 10^{-4} & 5\times 10^{-5}\\
5\times 10^{-5} & 3.5\times 10^{-4}\end{pmatrix}}
\def\Hmattot{\begin{pmatrix} 6.6\times 10^{-4} & -9.7\times 10^{-05}\\
-9.7\times 10^{-05} & 8.1\times 10^{-4}\end{pmatrix}}
\def\Hmatlinked{\begin{pmatrix} 8.3\times 10^{-4} & 9.6\times 10^{-4}\\
9.6\times 10^{-4} & 0.0025\end{pmatrix}}
\def\Hmatunlinked{\begin{pmatrix} 2.9\times 10^{-4} & -1\times 10^{-06}\\
-1\times 10^{-06} & 2\times 10^{-4}\end{pmatrix}}

\def\pvalWcentroidother{p = 0.022}
\def\pvalWcentrepointother{p = 0.0085}
\def\pvalWmodeother{p = 0.42}
\def\pvalWmodeWother{p = 0.89}
\def\pvallinkedmodeWother{p = 0.03}
\def\pvalunlinkedmodeWother{p = 0.57}
\def\pvalWmodeoriginal{p = 0.031}
\def\pvalwithOutsidecentroidoriginal{p = 0.2}
\def\pvalwithOutsidecentrepointoriginal{p = 0.012}
\def\pvalweiboXuJanmodeWother{p = 0.0075}
\def\pvalWHSMcentre{p < 5\times 10^{-4}}
\def\pvalWHSMcentreworldpop{p < 5\times 10^{-4}}
\def\pvalWHSMfirst{p = 8.9\times 10^{-5}}
\def\pvalWHSMfirstworldpop{p = 8.8\times 10^{-5}}
\def\pvalWHSMfirstWeibo{p = 10^{-6}}
\def\pvalWHSMcentreWeibo{p < 5\times 10^{-4}}
\def\pvalWHSMcentre{p < 5\times 10^{-4}}
\def\pvalWHSMcentreworldpop{p < 5\times 10^{-4}}
\def\pvalWHSMfirst{p = 8.9\times 10^{-5}}
\def\pvalWHSMfirstworldpop{p = 8.8\times 10^{-5}}
\def\pvalWHSMfirstWeibo{p < 0.0023}
\def\pvalWHSMcentreWeibo{p < 5\times 10^{-4}}

\def\pvalcentroidfifty{p < 5\times 10^{-4}}
\def\pvalcentrepointfifty{p = 0.001}
\def\pvalmodefifty{p = 0.74}
\def\pvalmodeWfifty{p = 0.12}
\def\pvalmodefivehundred{p = 0.001}

\def\distWMHSM{99~m}
\def\distWMCDC{341~m}
\def\distWeiboXuJanHSM{2929~m}

% Figure commands
\def \wpic {8cm}
\def \colspace {1pt}

% Define command to center figure larger than text width
% https://tex.stackexchange.com/questions/16582/center-figure-that-is-wider-than-textwidth
\makeatletter
\newcommand*{\centerfloat}{%
  \parindent \z@
  \leftskip \z@ \@plus 1fil \@minus \textwidth
  \rightskip\leftskip
  \parfillskip \z@skip}
\makeatother

%\linenumbers

\section*{Abstract}
The centrality of Wuhan's Huanan market in maps of December 2019 \Covid{} case residential locations, established by \citet{Worobey2022Science}, has recently been challenged by \citet[][\SC]{StoyanChiu2024}. \SC{} proposed a statistical test based on the premise that the measure of central tendency (hereafter, ``centre'') of a sample of case locations must coincide with the \textit{exact} point from which local transmission began. Here we show that this premise is erroneous. \SC{} put forward two alternative centres (centroid and mode) to the centre-point which was used by \citeauthor{Worobey2022Science} for some analyses, and proposed a bootstrapping method, based on their premise, to test whether a particular location is consistent with it being the point source of transmission. We show that \SC's concerns about the use of centre-points are inconsequential, and that use of centroids for these data is inadvisable. The mode is an appropriate, even optimal, choice as centre; however, contrary to \SC's results, we demonstrate that with proper implementation of their methods, the mode falls at the entrance of a parking lot at the market itself, and the $95\%$ confidence region around the mode includes the market. Thus, the market cannot be rejected as central even by \SC's overly stringent statistical test. Our results directly contradict \SC's and -- together with myriad additional lines of evidence overlooked by \SC, including crucial epidemiological information -- point to the Huanan market as the early epicentre of the \Covid{} pandemic.

\section{Introduction}

While the origin of the \Covid{} pandemic remains widely debated in the media and the wider public sphere, scientific contributions have established the central role played by Wuhan's Huanan Seafood Wholesale Market (hereafter ``Huanan market'') during the early days of the \Covid{} pandemic, given data currently available \citep{Worobey2021Science, Worobey2022Science, JiangWang2022, ACC2023bioRxiv}. Although this market almost immediately became the main focus of attention because so many of the very first identified \Covid{} cases worked or spent time there, the identification of other early cases without a known exposure risk at the market  \citep{Huang2020Lancet} introduced uncertainty about the role of the Huanan market as the potential source of the outbreak: to some, these cases appeared to be strong evidence that the epidemic in Wuhan began elsewhere and was perhaps already geographically widespread in the city in December 2019 \citep{Cohen2020ScienceHSM}.

An important turning point in the scientific understanding of the pandemic's origin occurred in 2021, with the release of a report from a joint World Health Organization-China \Covid{} origins study, the ``WHO-convened global study of origins of SARS-CoV-2: China part'' (hereafter ``WHO-China report'') \citep{WHO2021}. This report contained maps of residential locations of early \Covid{} cases in Wuhan, those whose symptoms had begun in December 2019 (hereafter, ``\ecl''; there are no earlier known onset dates). The study report noted that the residences of these cases were concentrated in the central districts of Wuhan, where the Huanan market was also located.\footnote{\quotestyle{There was a concentration of cases, both laboratory-confirmed and clinically diagnosed, in the central districts (which include the Huanan market). The earliest cases were mostly resident in the central districts of Wuhan, but cases began to appear in all districts of Wuhan in mid-to late December 2019}; \citet[][p.~44]{WHO2021}.}

Starting a few months later, \citet{Holmes2021Cell} and \citet{Worobey2022Zenodo, Worobey2022Science}  manually extracted coordinates from the maps of the WHO-China report and conducted spatial analyses. \Citet{Worobey2022Science} established that the early cases were geographically centred near the Huanan market to a degree that would be extremely unlikely unless the outbreak had begun there or at another location near it. Based not just on these spatial findings but also many other lines of evidence, \citet{Worobey2022Science} concluded that the Huanan market was the \quotestyle{early epicenter} of the \Covid{} pandemic.

Of the geographic findings in \citet{Worobey2022Science}, the most striking and consequential involved the early cases epidemiologically unlinked to the market (i.e., those who had not been to the market and who reported no known contact with anyone who had). (Hereafter, as in \citet{Worobey2022Science}, we use ``linked cases'' and ``unlinked cases'' only to refer to ascertained epidemiological linkage, regardless of whether cases were geographically associated with the market.) Analysed separately from linked cases, who worked at or were otherwise knowingly connected to the market, not only was the geographical centre of the unlinked case locations revealed to be closer to the market than expected from a null population distribution, but it was also closer to the Huanan market than the centre of linked case locations \citep{Worobey2022Science}. It had previously been argued that the initial requirement of a market link for case definition could have biased the early case data towards the Huanan market. 
While it is important to take into account potential ascertainment bias, multiple arguments indicate that it was more limited than some claimed. First, case definitions were rapidly updated and did not include a market link \citep[][Annex E3]{WHO2021}. Second, a large fraction of the cases with symptom onset in December were identified retrospectively, after case definitions had been updated. Third, the very existence of large numbers of unlinked December cases -- about twice as many as linked cases -- directly disproves the claim that the cases with onset in December 2019 were identified via their link to the market. Fourth, the spatial structure of these early, unlinked cases provided compelling evidence for a connection between the Huanan market and the onset of the outbreak in Wuhan \citep{Worobey2022Science}.

\citet[][hereafter, \SC]{StoyanChiu2024} contested some of the results presented by \citet[][hereafter, \WHSM]{Worobey2022Science}. \SC's criticisms, however, stem from \begin{paraenum} \item profound misunderstandings of \WHSM's study and objectives, \item a crude technical implementation of a key function, and \item a very narrow scientific perspective. \end{paraenum}

\section{The source of an outbreak is not expected to be at the \textit{exact} centre of case locations\label{sec:SCcentre}}

\subsection{A profound misunderstanding of \citet{Worobey2022Science}'s claims}

While \WHSM{} inferred, based on both geographical \textit{and} other evidence, that the market was the \quotestyle{epicenter} of the pandemic, they never claimed that the market was positioned \textit{exactly} at the centre of the cloud of the \ecl.

\begin{figure}
\centerfloat
\begin{tabular}{ll}
$(a)$ December 2019 cases ($n = \ncases$) & $(b)$ Unlinked December 2019 cases only ($n = \nunlinked$) \\
\includegraphics[width = \wpic]{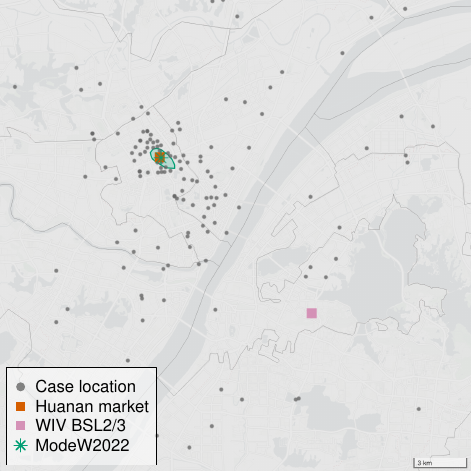}
&
\includegraphics[width = \wpic]{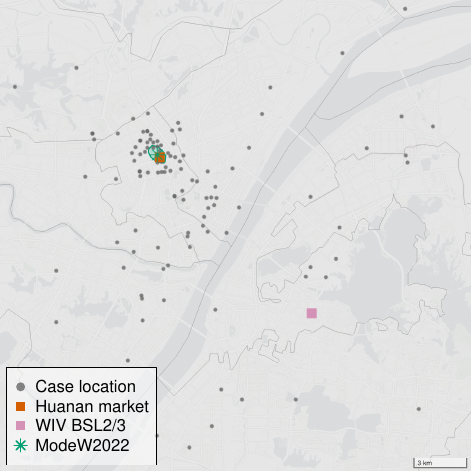}\\
% Weibo
$(c)$ January 2020 Weibo cases ($n = \nWeiboXuJan$) & 
$(d)$ February 2020 Weibo cases ($n = \nWeiboXuFeb$) \\
\includegraphics[width = \wpic]{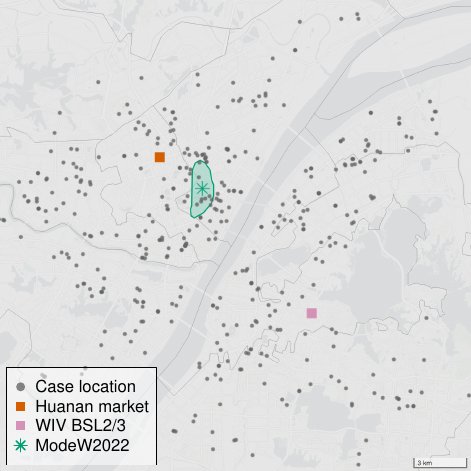}
&
\includegraphics[width = \wpic]{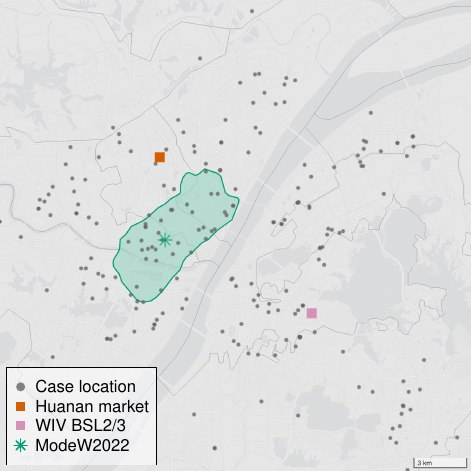}
\end{tabular}
\caption{Positions of the modes of case locations and the associated $95\%$ confidence surfaces, inferred by computing the centres of bootstrap pseudoreplicates, using \citet{Worobey2022Science}'s KDE function. $(a)$ all \ecl{} from \WHSM; $(b)$ unlinked cases only; $(c)$ Weibo cases with confirmation time in January 2020; $(d)$ Weibo cases with confirmation time in February 2020. The \ecl{} ($(a)$--$(b)$) were compiled by \WHSM{} from \citet{WHO2021}; the Weibo data ($(c)$--$(d)$) were compiled by \citet{Xu2020Weibo}.}
\label{fig:WuhanMode}
\end{figure}

Even in cases where a contagious disease is initially transmitted from a point source, there are many reasons why that source will not fall exactly at the centre of case residential locations. 
Geographic constraints, like the arrangement of streets or unequal population densities in different directions from the source, are obvious reasons. Long distance dispersal events as people move around an urban landscape is another. 
There is also not a single possible case map, and therefore not a single centre of case locations: workplaces could for instance have been mapped instead of residences -- there would have been many more points right at the Huanan market. The case map would also have been different if the locations of all individuals at a precise time on a precise day had been mapped. 
In addition, as time goes by, the distribution of cases will change, and not necessarily isotropically. Indeed, as expected as neighbourhood transmission gives way to widespread transmission across a city, in Wuhan the distribution shifted dramatically over time from the neighbourhoods around Huanan market towards the centre of the city’s high density areas, southeast of the market \citep[see Figs.~1D--E in][and our Figure~\ref{fig:WuhanMode}]{Worobey2022Science}. This illustrates that even if an outbreak is initially clustered around its point source, time can be the biggest enemy to detecting that signal. 

For these reasons, the rejection of \SC's null hypothesis of exact centrality would not actually tell us much about the role of the Huanan market as the possible source. It is therefore preposterous to exclude the market on the basis of a too-stringent definition of ``centre''.

\subsection{Impact on the choice of statistical tests}

An important divergence between \WHSM{} and \SC{} relates to the question of how best to test the hypothesis that a site -- identified in advance of spatial analyses by compelling non-geographic evidence -- may have been the place from which the early outbreak spread. For \Covid{}, the Huanan market is the preeminent such site. 

\WHSM{} asked whether the centre of the \ecl{} was \textit{closer to} the Huanan market than other possible ``epicenter'' locations in the city were. \SC{} asked whether \ecl{} were \textit{centred on} the Huanan market. As detailed already, there is no reason to expect that the centre of early case locations will necessarily be exactly centred on the initial source. \SC{}'s null hypothesis is therefore not appropriate, and rejecting it is not informative. \WHSM's tests, on the other hand, compared the distribution of \ecl{} to chosen null distributions, and they characterized the distributions through their proximity to the Huanan market. 

\WHSM{} considered two types of null distributions. The first was Wuhan's population, using age-adjusted data from \worldpop. This null distribution assumes that the disease has had so much time to spread that any signal of its initial source has vanished, and the distribution of cases matches the underlying distribution of the population of susceptible hosts. However, population density data for Wuhan from \worldpop{} may not be as accurate as other available sources,\footnote{Daniel A. Walker, 2023-08-23, \quotestyle{The worldpop data are highly discordant with other sources regarding Wuhan population density}, https://archive.is/WVSDv.} and we therefore use another source in our analysis here \citep[][shown in Figure~\ref{fig:popdens}]{Peng2020Pop}. 

The second type of null distribution considered by \WHSM{} consisted of self-reported \Covid{} cases on a social media service (Weibo), with onsets in Jan-Feb 2020 \citep[from ][]{Peng2020IJGI}. This distribution provided a snapshot of case distributions later in the spread of the disease, but from a point when clustering of cases may still have existed. In our analysis, we use another dataset from the same original Weibo source \citep{Xu2020Weibo}, for which onset dates were available, unlike for the Weibo dataset used in \WHSM{} (see Figure~\ref{fig:WuhanMode}c; we use the January data as null distribution). 

In their first set of tests, \WHSM{} first compared the median distance to Huanan of sets of locations drawn from a null distribution, to the median distance to Huanan of the \ecl{}. Tests presented in \WHSM's main text were based on a null distribution defined using \worldpop{} population density data, and as such, did not include a notion of clustering: again, these were tests of the scenario outlined above, that the outbreak had had enough time to spread such that the spatial distribution of cases matched that of the population of hosts, and therefore lacked clustering. Although this is not a hypothesis that \WHSM{} thought was likely to be true, in part because it is refuted by genomic evidence indicating that the Wuhan outbreak likely began only in late November \citep{Pekar2022Science}, the idea of a hidden outbreak that long predated cases at the Huanan market has been proposed by others \citep[e.g.,][]{Cohen2020ScienceHSM}. Despite the limitations of this first test conducted by \WHSM, the rejection of this null hypothesis thus provided an important confirmation that this was not the case. Finally, unlike the population density data, and contrary to \SC's assertions, the second null distribution considered by \WHSM{}, Weibo case data, did encompass case clustering and human-to-human transmission  \citep[see][Table~S4, ``median distance'']{Worobey2022Science}. 

\begin{figure}
\centerfloat
\includegraphics[]{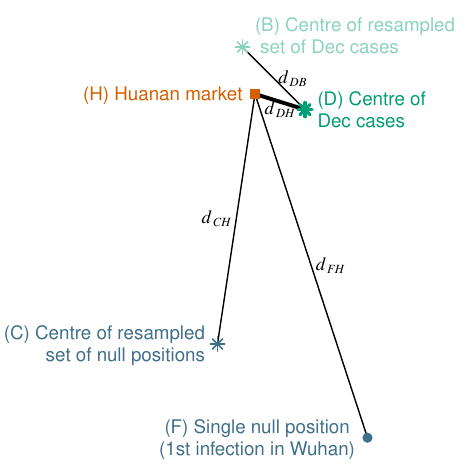}
\caption{Schematic of the different distances compared in the various tests of market closeness and centrality. }
\label{fig:schematic}
\end{figure}

Case clustering was also encompassed in the second, and more important, set of tests conducted by \WHSM. These tests compared the distance from the centre of the \ecl{} to the market against the distances to the Huanan market of single points drawn from a null distribution of plausible starting points of the outbreak. 
On Figure~\ref{fig:schematic}, this amounts to comparing distance $d_{DH}$ to the distribution of distances $d_{FH}$. Implicit in this test was the idea that the single point drawn in each replicate represented the first human infection in Wuhan, which then acted as the ultimate source of all subsequent infections, and that the centre of the resulting cluster of cases would have been near the position of the first infection. (It was assumed, but not explicitly modelled, that simulating a clustered outbreak around each first infection, then inferring its centre, would have produced nearly identical null distributions, just with some jitter around the location of each first infection.) 
For both null distributions considered by \WHSM{} (\worldpop{} population density, and Weibo cases), the centre of the \ecl{} was significantly closer to the Huanan market than were the putative centres of case locations drawn from each null distribution \citep[see][Table~S4, ``center-point distance'']{Worobey2022Science}. As part of the current study, we reproduced this test and its results with different data sources for the two null distributions considered by \WHSM, and using the mode as centre (see below for a description of the mode). 
The mode of the \ecl{} was significantly closer to the Huanan market, whether we compared it to first infection locations drawn from Wuhan's population density ($\pvalWHSMfirst$; $\nrepsfirstinf$~draws), or from the January 2020 Weibo cases ($\pvalWHSMfirstWeibo$).

We also implemented a version of the test in which we compared the distance between the market and the centre of the \ecl{} ($d_{DH}$ in Figure~\ref{fig:schematic}) against the distances between the market and the centres of sets of the same number of cases ($n = \ncases$) drawn from a null distribution (i.e., the distribution of distances $d_{CH}$ in Figure~\ref{fig:schematic}). We used the mode as centre. This test is meant to provide a more apples-to-apples comparison than the previous ones described above, by comparing the positions of putative centres inferred from the null distribution of cases with the centre of the observed cases. It is additionally meant to mitigate the clustering issue -- the distribution of the centres of sets of points is much more condensed than the distribution from which these points are drawn (see Figure~\ref{fig:modesRandom}).  
We found that the centre of the \ecl{} was significantly closer to the Huanan market than were the centres of randomly sampled cases, whether we used used sets of ($n = \ncases$) locations from Wuhan's population density as the null distribution ($\pvalWHSMcentre$; $\nreps$ replicates; Figure~\ref{fig:modesRandom}a) or from the January 2020 Weibo cases ($\pvalWHSMcentreWeibo$; $\nreps$ replicates; Figure~\ref{fig:modesRandom}b).

While \WHSM{} asked whether the centre of the \ecl{} was \textit{closer to} the market than (appropriate) random locations, \SC{} asked a different question: whether the \ecl{} were \textit{centred on} the market. \SC's question did not involve the identification of a null distribution; instead, \SC{} bootstrapped \ecl. They compared the distance between the Huanan market and the centre of the \ecl{} ($d_{DH}$ in Figure~\ref{fig:schematic}), against the distances between centres of bootstrapped sets of \ecl{} and the centre of the \ecl{} ($d_{DB}$ in Figure~\ref{fig:schematic}). Their tests (and in particular, as we will see below, their method for computing the mode) led to the rejection of the hypothesis that the market was the centre of the \ecl, because the market was just outside of the region covered by the bootstrapped centres. Again, though, the source of the outbreak is not expected to be exactly at the centre of the \ecl{}; \SC's comparison is therefore based on a false premise. We nevertheless explore \SC's approach to further highlight its shortcomings.

\section{Defining an appropriate centre of case locations and computing its location}

\SC{} discussed different types of centres of a cloud of points. They asserted that a centre-point, defined by \WHSM{} as the coordinate-wise median, is \quotestyle{a questionable choice}, because of its lack of rotational invariance. \SC{} proposed two alternative centres, which are rotationally invariant: the centroid (defined as the coordinate-wise mean) and the mode (defined as the location of the maximum of a corresponding spatial density function). We will discuss the relevance of the three types of proposed centres, and will show that \SC's rejection of the Huanan market as centre was due, specifically, to a crude implementation of a kernel density estimation (KDE).

\subsection{The impact of the lack of rotation-invariance of the centre-point is limited}

A rotation-invariant centre does not depend on the orientation of the axes, which, all else being equal, seems to be a desirable property for the centre of a cloud of points -- while natural, the latitude-longitude orientation is not the only orientation of axes one could use to reference locations. We tested how the position of the centre-point changes with the orientation of the axes. We find that, while the computed position of the centre-point does indeed change slightly, it remains in a circumscribed area, in the vicinity of the Huanan market (Figure~\ref{fig:rotationcentre-point}). The centre-point may not be strictly speaking rotation-invariant, but the overall conclusions of \WHSM{} are not affected. 

\subsection{A centroid is affected by extreme values}

\SC{} proposed to use a centroid (coordinate-wise means) as the first of two rotationally invariant alternatives to a centre-point. \WHSM{} had however chosen a centre-point because a median is more robust to extreme values than a mean. The \ecl{} data used by \WHSM{} contain extreme values, and are not symmetric (Figure~\ref{fig:distributionslonlat}), so a centre based on medians was preferred by \WHSM.\footnote{\quotestyle{We used medians rather than means for our analyses so as
to not give undue influence to outliers like those that can be seen in fig. S8.}, \citet[][Supplementary text]{Worobey2022Science}.}

In addition, because their focus was on Wuhan, the clear epicentre city of the pandemic, \WHSM's data did not include ten additional cases mentioned in the China-WHO report \citep[][Fig. 23, p.45; annex Fig.~3, p.147]{WHO2021}. These cases were in seven cities in Hubei province, but outside of Wuhan, and their features (sequenced or not, type of case confirmation) were not described. To investigate the impact of minor changes to data on centroids and centre-points, we digitised the positions of these cities, and added them to the \ecl{} data from \WHSM{} (Figure~\ref{fig:NewCasesZoomOut}). We then repeated \SC's methodology and computed the new positions of the centroid and the centre-point, and their $95\%$ confidence regions using \SC's bootstrap methodology (Figure~\ref{fig:WuhanNewCases}; $\nreps$ bootstrap resamples). As expected since it is a median-based centre, the centre-point and its $95\%$ confidence surface are barely affected by the inclusion of the seven Hubei cases (Figure~\ref{fig:WuhanNewCases}a--b). However, because the new cases are distant from the centre of Wuhan (about 150 km for the furthest away), their inclusion changes the position of the centroid, but also considerably widens the $95\%$ confidence surface (Figure~\ref{fig:WuhanNewCases}c--d). While the market was outside of the confidence region when only cases within Wuhan were considered, the market is now clearly inside when the seven Hubei locations are added ($\pvalwithOutsidecentroidoriginal$), and its centrality is not rejected anymore by \SC's (albeit dubious) test.

\Citet{Worobey2022Science} chose centre-points to avoid just such issues with extreme values. 
Indeed, it is not advisable to rely upon a measure of central tendency that is, like the centroid, so sensitive to minor changes to the data.

\subsection{Stoyan and Chiu's exclusion of the Huanan market is due to the use of a circular kernel with too high a bandwidth}

The mode, i.e. the peak of the underlying spatial density distribution, was the second alternative centre proposed by \SC. This suggestion, described as rotation-invariant (unlike the centre-point), is (like the centre-point but unlike the centroid) relatively immune to extreme values. As such, in important ways it combines the main advantages of both the centre-point and the centroid. 
The determination of a mode, however, requires the choice of a kernel function and, specifically, of a bandwidth for kernel density estimation.

\SC{} used a circular Gaussian kernel and a large value for their bandwidth, $3000$~m, which had been considered in the preprint version of \citet{Worobey2022Zenodo}, in which kernel density estimates (KDE) were computed using ArcGIS Online. The published peer-reviewed article \citep{Worobey2022Science}, however, employed different software and methods than the preprint (\texttt{R}, function \texttt{kde} from the \texttt{ks} package \citep{ksPackage}), and, importantly, automatic bandwidth selection, yielding a matrix (via the \texttt{Hpi} function) instead of a scalar (Figure~\ref{fig:Wuhan_bandwidth} illustrates the impacts of these choices on the shape of the spatial density function). Rather than attempting to replicate the methods in the published \textit{Science} paper, \SC{} emulated (partially) the \citet{Worobey2022Zenodo} preprint methodology, using a circular kernel (with the \texttt{bkde2D} function from the \texttt{KernSmooth} package \citep{KernSmoothPackage}, and setting the bandwidth to $h = 3000$~m).

Implementing \SC's bootstrap test but with \WHSM's more sophisticated KDE function, we find that the mode lies at the entrance of a parking area at the Huanan market, and that the market is within the mode's $95\%$ confidence region (Figure~\ref{fig:WuhanMode}a;  $\pvalWmodeWother$; and see Figure~\ref{fig:WuhanZoomMarketMode} for a detailed view). The market is thus in fact not rejected as the centre of the distribution of the \ecl, even using \SC's preferred but inappropriately stringent hypothesis test. A similar non-rejection is obtained with Stoyan and Chiu (2024)'s KDE function when using a bandwidth automatically determined using a principled rather than arbitrary approach ($h = \hall$; Figure~\ref{fig:WuhanModeSC}b; $\pvalWmodeother$). 

\section{Stoyan and Chiu overlooked key epidemiological data}

\subsection{A limited perspective}

A key feature of \SC's analysis is its limited perspective. Indeed, the authors overlooked almost the entirety of the scientific evidence available about the early period of the pandemic, considering only the \ecl{}. Other pieces of evidence are however the reason that the Huanan market has, since almost the first inklings that a new disease may have emerged, been the prime suspect as the source of Wuhan's \Covid{} outbreak. The link to the Huanan market of early patients with pneumonias of unknown aetiology cases that would turn out to be \Covid{} helped lead to the discovery of the new disease, and the market was rapidly closed down \citep{Li2020CDC, Yang2024book}; this single market accounted for a substantial proportion of the earliest-onset \Covid{} patients in what is a very large city \citep{Worobey2021Science}; it had been identified years before the pandemic as a site where viruses with pandemic potential might jump from animals into humans \citep{ZhangHolmes2020}; and it was one of only four markets in Wuhan with sustained sales of live mammals from intermediate host species known to harbour SARS-related coronavirus \citep{Xiao2021SciRep}. Additionally, although the data were not public until after \citeauthor{Worobey2022Science}'s study was conceived and published, genetic traces of these animals were found in environmental samples from the market, in stalls where \sct{} was also detected \citep{Liu2023Nature, ACC2023Zenodo, ACC2023bioRxiv, Debarre2024VE}. 

\subsection{Linked vs.\ unlinked cases}
Notably, \SC{} also overlooked the fundamental difference between the linked and the unlinked December 2019 cases as it relates to spatial analyses. Cases epidemiologically linked to the market worked there or had shopped there, and for some were clearly involved in human-to-human transmission chains within the market \citep{Li2020NEJM}. These linked cases are thus expected to be found around the Huanan market only because people often shop close to where they live, and often live not too far from where they work. Unlinked cases, on the other hand, are only expected to be found to reside near the market if local transmission originated from the market, and only at an early point in the outbreak, before the virus had spread widely. 
The observed spatial pattern for unlinked cases was therefore particularly noteworthy. And yet, even though information on market links was available in the case dataset used by \SC, they did not use it in their analyses. With \citet{StoyanChiu2024}'s methodology and \WHSM's KDE function, we find that analysing only unlinked cases gives the same results as considering all cases regardless of linkage: the Huanan market cannot be rejected as being at the very centre of the unlinked cases (Figure~\ref{fig:WuhanMode}b; $\pvalunlinkedmodeWother$). 

\subsection{Missing the tree for the forest}

In their paper, \SC{} proffer multiple other landmarks, in the vicinity of the Huanan market, that could have played a role in the origin of the pandemic, based on the argument that they are close to one of the centres they computed from the \ecl. It is of course true that, when considering the \ecl{} data in an epistemic vacuum, it is not possible to rule out alternative sources also in the vicinity of their centre. Other data, however, help assess the plausibility of the other landmarks proposed by \SC. This is for instance the case for a complex called ``Wanda plaza''.\footnote{\label{foot:WandaLocation}We note that this complex, as well as other landmarks, were improperly located by \SC. \SC{} seem to have used Google Maps to extract coordinates, but did not take into account the offset on maps of China shown on Google Maps. We present corrected locations (see Figure~\ref{fig:landmarks}). We also note that the complex does not just include a shopping centre, but also residences. The social media check-in data used by \SC{} do not seem to differentiate between the different uses of the place.} 
\SC's \textit{post hoc} reasoning, in inferring a centre of the \ecl{} and then looking for a landmark near it, completely misses the fact that human-to-human transmission did happen in relation to the Huanan market \citep{Li2020NEJM}. The market did play an epidemiological role, and is of interest not just because of its position in Wuhan: it was of interest because of its role in the early days of \Covid, before it was known that the centre of the \ecl{} would be revealed to be so close to it -- as already described above. 
Landmarks are to be chosen by taking into account actual potential sources. While the market is one such potential source, Wanda Plaza has, to our knowledge, never been mentioned as such until \SC. Highlighting the existence of other data linking cases to the Huanan market, and the absence of such evidence for other landmarks in Wuhan, was actually the point of the ``No other location except the Huanan market clearly epidemiologically linked to early COVID-19 cases''  paragraph in \WHSM's supplementary text, that \SC{} seem to have misunderstood.

\SC{} used social media check-in data to argue that Wanda Plaza was a much more visited place than the Huanan market. Putting aside the fact that Wanda Plaza is more than just a shopping centre (see footnote~\ref{foot:WandaLocation}), \SC's argument reveals a profound misunderstanding of the use of social media check-in data by \WHSM. The Huanan market being much less visited than Wanda Plaza actually emphasises the extraordinary role played by the Huanan market in the early days of \Covid. The market was not a hotspot in Wuhan in terms of visitation check-ins -- it was one of hundreds of possible early sites for case clustering, and even amongst markets across the city, there were 70 with (up to hundred times) more check-ins \citep{Worobey2022Science}. And yet, the Huanan market was the hotspot for December cases and in the distribution of the \ecl. 

A new location of the Wuhan Centre for Disease control (WCDC; see Figure~\ref{fig:alllandmarks}) was also proposed as the potential source of the Wuhan outbreak. WCDC was the focus of one of the earliest speculations about a potential non-natural origin of \sct{} \citep{Xiao2020RG}, and was identified because of its proximity to the market (i.e., \textit{post-hoc}). However, WCDC had moved to its new location close to the market on December 2nd, 2019 \citep[][p.122 and Annex D5]{WHO2021}, leaving little time or no time for its BSL2 laboratory to be operational -- and this laboratory did not conduct elaborate virology experiments like the Wuhan Institute of Virology. While WCDC had been involved in wildlife sampling, no mammalian samples were moved between the old and new locations \citep{Holmes2024}. Finally, as noted in \citet{Worobey2022Zenodo, Worobey2022Science}, only one WCDC staff member had a positive serology result \citep[][Annex D5]{WHO2021}, to be contrasted with the at least $30$ Huanan market vendors who were identified as cases. Moreover, this individual was reported to have been infected by a family member and not at WCDC \citep[][Annex D5]{WHO2021}.

\begin{figure}[h!]
\centerfloat
\includegraphics[width = \wpic]{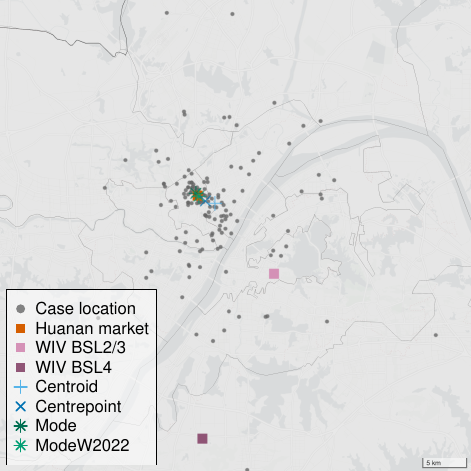} 
\caption{Zooming out. The two campuses of the Wuhan Institute of Virology (WIV) are shown. }
\label{fig:WuhanZoomOut}
\end{figure}

Finally, it is useful to zoom out and to consider the distribution of the \ecl{} in Wuhan relative to the locations of the campuses of the Wuhan Institute of Virology (WIV). This institute has been at the centre of speculations around a potential lab leak since early 2020. However, Figure~\ref{fig:WuhanZoomOut} shows the clear disconnect between the \ecl{} and the WIV campuses.

\section{Discussion}

Using the same methodology as \citet[][(\SC)]{StoyanChiu2024} on the data extracted by \citet[][(\WHSM)]{Worobey2022Science}, but with more appropriate parameters and implementation, we arrive at opposite conclusions, and show that Wuhan's Huanan market cannot, after all, be rejected as the source of the \Covid{} pandemic -- even with an overly-stringent test. 

This result is not surprising given the amount of other evidence pointing to the Huanan market. In their article, \SC{} dismissed these other pieces of information, characterising them as \quotestyle{some non-statistical argument}. The question of the origin of \sct{}, however, requires the consideration of multiple lines of evidence -- and these include other statistical arguments. 
Epidemiological links of numerous early cases to the market were identified early on \citep{Worobey2021Science, Yang2024book}, and our study re-emphasizes the spatial proximity between the market and \ecl{}, including cases for which no link to the market had been identified \citep{Worobey2022Science}. 

A second main line of evidence involves wildlife sales: animals susceptible to \sct{} were demonstrated to have been in the Huanan market \citep{Xiao2021SciRep, ACC2023Zenodo, ACC2023bioRxiv, Liu2023Nature, Debarre2024VE}, 
which was the most active of only four markets in Wuhan with consistent sales of live mammals that are the most plausible intermediate hosts of \sct{} \citep{Xiao2021SciRep}. In addition, there was a high concentration of \sct-positive samples in the corner of the market where most live wildlife was sold \citep{Wu2020CCDC, Worobey2022Science, ACC2023bioRxiv}, and genetic traces of \sct{} and animals were detected in the same stalls and in drains directly below and downstream of these stalls \citep{Liu2023Nature}. 

A third line of evidence involves \sct's genetic diversity. There were two early \sct{} lineages, A and B. Sequenced market-linked cases were of lineage B, but the earliest-known lineage A cases were close to the market \citep{Lu2020, Worobey2022Science}, and both lineages A and B were detected in environmental samples at the market \citep{Liu2023Nature}. In addition, statistical analyses of the \sct{} molecular clock strongly reject a single introduction into humans of a bat virus-like lineage A-haplotype ancestor from which lineage B evolved within humans \citep{Pekar2022Science}. The inconsistency between ancestor inference with molecular clock methods (rejecting lineage A) vs.\ with the consideration of related sarbecoviruses (favouring lineage A), led to the suggestion that there may not have been only a single introduction of the progenitor virus into humans, but multiple ones \citep{Pekar2022Science} -- a scenario that is fully expected at a wildlife market, but not with a lab origin. Importantly, a market origin is not contingent on multiple spillovers, and any origin scenario needs to account for the presence of early lineage A and lineage B both in and near the market. 

Finally, temporal arguments also align with a market origin: phylogenetic and epidemiological analyses suggested that the \Covid{} outbreak in Wuhan began only shortly before the earliest known cases at the market became symptomatic \citep{Jijon2023, Pekar2022Science}. Consistent with this, serological testing of stored blood samples, collected from September to December 2019 from more than 32,000 individuals in Wuhan, revealed that not a single one had neutralizing antibodies against \sct{} \citep{Chang2023}, i.e., that \sct{} was not widely circulating in Wuhan in the last quarter of 2019. And it was not until December 27th--29th 2019 that the first trickle of reports of \Covid{} cases was recognized by local health authorities as a concern \citep{Yang2024book}. All but one of the hospitals first reporting pneumonia of aetiology cases were very near the Huanan market \citep{Yang2024book}.

We agree with \SC{} that there are of course many ways that the spatial pattern of cases in an emerging epidemic of an infectious respiratory disease could have deviated from the Huanan-centred pattern observed here. Indeed, data of individuals self-reporting their \Covid{} infection on social media (Weibo) to get help, mostly in January and February 2020, indicate that by that time cases had spread widely across the city, with locations of these cases largely recapitulating patterns of population density, particularly for older people \citep[][and our Figures~\ref{fig:WuhanMode} and \ref{fig:popdens}]{Peng2020IJGI}. By then, the pattern placing the Huanan market at the centre of the earlier, December 2019-onset cases, had been obscured by subsequent spread \citep{Worobey2022Science}.

Even in cases of non-transmissible diseases, a source may not be located at the centre of a cloud of case locations, when environmental factors come into play. For instance, the anthrax cases near the Sverdlovsk military facility in Spring~1979 were to the south-east of the building, reflecting the direction of the wind \citep{Meselson1994Science}.
 
\begin{figure}[h]
\centerfloat
\begin{tabular}{ll}
$(a)$ Mode, \SC's KDE, $h=500$ & $(b)$ Mode, \WHSM's KDE \\
\includegraphics[width = \wpic]{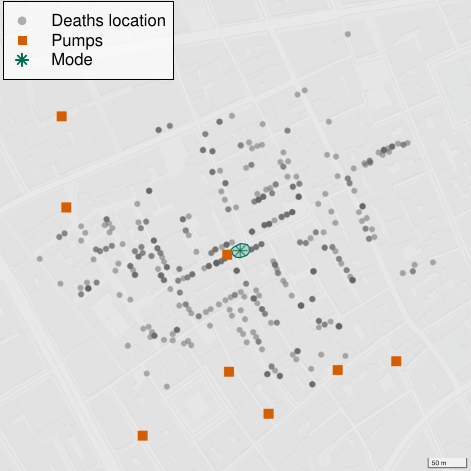} &
\includegraphics[width = \wpic]{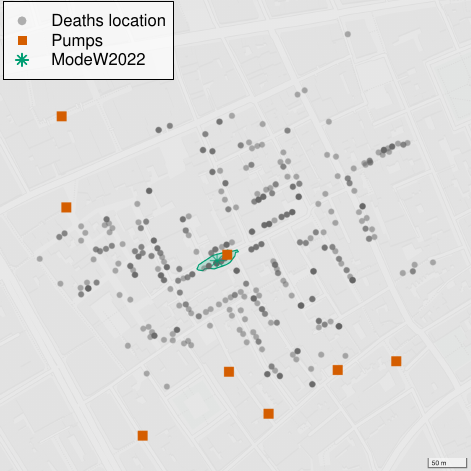} 
\end{tabular}
\caption{Identification (or not) of the Broad street pump using John Snow's data for the 1854 London cholera outbreak, using \citet{StoyanChiu2024}'s bootstrap method (the background map shows modern London). The panels show results for different centres and KDE functions. $p$ values: 
$(a)$ Mode with \SC's KDE and $h = 500$, $\pvalmodefivehundred$; $(b)$ Mode with \WHSM's KDE $\pvalmodeWfifty$.}
\label{fig:snow}
\end{figure}

Conversely, in the classic case of `John Snow and the Broad Street pump', which established the field of spatial epidemiology, the Broad street pump was a well-identified source during the 1854 London cholera outbreak, with quite a clearly central position compared to case fatality locations \citep{Snow1855, Falcone2020}. As with live wildlife markets and \Covid, water pumps were pre-identified by John Snow as potential point sources of water-borne disease. Snow did not go for a pump, of all possible features in the landscape of the city, at random; he focused on pumps because he had other reasons to suspect a role of water in spreading cholera. But even then, inappropriate criteria of the sort employed by \SC{} would exclude the pump as the source. Indeed, we applied \SC's methodology and their KDE function on the Snow data; the pump was outside of the $95\%$ confidence surfaces (see Figure~\ref{fig:snow}a) and thus rejected as the centre of that cholera outbreak. Yet we all understand that the source of the cholera outbreak was not a cobblestone near the pump, nor a building across the street from the pump, but the contaminated water flowing from the pump itself. When we used \WHSM's KDE on the Snow data on the other hand, the pump was not rejected as central (Figure~\ref{fig:snow}b). 

Real-life data are always finite. However, while data on the early days of \Covid{} could have been shared more quickly, in more accessible formats, and while some data are known to exist but have not been shared yet \citep{Holmes2024}, the amount of information available is unprecedented for the emergence of a new disease. Data from various sources and of various types, all indicate that wildlife sales in the Huanan market played a key role in the onset of the \Covid{} pandemic. 

In this way, the Huanan market hypothesis is fundamentally distinct from lab leak hypotheses, which were important to consider \citep{Bloom2021Science}, but which encompass many unsupported, highly unlikely, and mutually exclusive scenarios. (For example, if the virus escaped from the Wuhan CDC, then it did not escape from the Wuhan Institute of Virology; if it came from a mine in southern China then it did not come from Laos; if this was a fieldwork accident then it was not an engineered virus, etc.) The Huanan market hypothesis, in contrast, has given rise to multiple predictions that were later confirmed as additional evidence arose. These include \begin{paraenum} \item that the \sct{} lineage A, found to be geographically associated with the Huanan market \citep{Lu2020, Worobey2021Science, Worobey2022Science}, would eventually be found to have been at the market \citep{Liu2023Nature}, and \item that cases unlinked epidemiologically to the market would be found to reside around the market \citep[and see][for other examples of such predictions]{Debarre2024VE}. \end{paraenum} 
Contrary to what \SC{} contended to have rebutted,\footnote{Their article's title reads \quotestyle{Statistics did not prove that the Huanan Seafood Wholesale Market was the early epicentre of the COVID-19 pandemic}.} \citet{Worobey2022Science}'s study did not claim to ``prove'' that the pandemic started at the market. Statistics do not offer ``proof'', but rather a means to think clearly about how probable different hypotheses are. Our results show that it is highly probable that community transmission in Wuhan began in the neighbourhoods directly surrounding the Huanan market. Other evidence shows that the Huanan market was deeply unlikely to be at the centre of the Wuhan outbreak by pure chance \citep[i.e., if the Wuhan outbreak had not begun there;][]{Worobey2022Science}. We should not lose sight of how remarkable it is that we have such unique insights into the origin of this pandemic, and we would do well to be mindful of the lesson all this evidence provides: that to prevent SARS-CoV-3 from emerging, we must reduce opportunities for pathogens with pandemic potential to be brought into the heart of large human populations via the live animals that harbour them.

\section*{Acknowledgements}
We thank Alex Crits-Christoph, Zach Hensel, Peter Jacobs, Josh Levy, Lorena Malpica, and Marc Suchard for discussions and/or comments. \\
This project has been funded in part with federal funds from the National Institute of Allergy and Infectious Diseases, National Institutes of Health (NIH), Department of Health and Human Services (contract no. 75N93021C00015 to M.W.)

\section*{Competing interests}

M.W. has received consulting fees from GLG and compensation for expert testimony from Endurance Specialty Insurance.
 
\section*{Data and methods}

We based our analyses on \citet{StoyanChiu2024}'s \texttt{R} script (provided as supplementary information) and on \citet{Worobey2022Science}'s data and code, shared on Zenodo (\url{https://zenodo.org/records/6908012}; \texttt{scripts/construct\_KDE\_contours.R}). The data we used and our \texttt{R} \citep{Rcitation} scripts are available on Zenodo at \url{https://doi.org/10.5281/zenodo.10779463}. The maps were drawn thanks to the \texttt{R} \texttt{leaflet} package \citep{leafletPackage}.

\subsection*{Data sources}

Like \citet{StoyanChiu2024}, we used case location data shared by \citet{Worobey2022Science}, which were manually extracted from figures in the China-WHO Joint mission report \citep{WHO2021} (see \citet{Worobey2022Zenodo, Worobey2022Science} for the extraction method). Following a similar methodology, we manually extracted the seven additional locations shown in annex Fig.~3 of the China-WHO Joint mission report \citep{WHO2021} (there are ten additional cases, but seven locations, and we do not know which location(s) had multiple cases; we therefore only placed seven additional points). 
Because some maps in the WHO-China report were too blurry, the linked or unlinked status of some cases is uncertain; the number of identified linked \ecl{} in \WHSM{} -- $\nlinked$ out of the $\ncases$ cases recovered from a map that contained $164$ cases within Wuhan \citep[][Annex E4]{WHO2021} -- is lower than the reported number of linked cases in the report -- $55$ out of $168$, including one or more Huanan-linked cases outside of Wuhan; the $168$ number excludes $6$ cases, out of the total of $174$ December 2019 cases, whose exposure history was unknown \citep[][Annex E4]{WHO2021}. \WHSM{} tested the robustness of their results to this (and other) sources of uncertainty \citep[][Supplementary information]{Worobey2022Science}.

The Snow data were from the dataset provided by \citet{Falcone2020}. The Weibo case data that we used were compiled by \citet{Xu2020Weibo}. 

Population density data were manually digitised from \citet[][Fig.~10]{Peng2020Pop} (This imperfect solution was chosen for lack of access to data). We first increased contrast between consecutive colors using \textsc{Gimp}, and then extracted the positions of squares with WebPlotDigitizer \citep{WebPlotDigitizer}. We georeferenced the figure using the positions of two landmarks (tips of islands). We programmatically re-aligned all positions.

\subsection*{Landmarks}

The position of \citet{StoyanChiu2024}'s additional landmarks was clearly erroneous (as revealed by a comparison of their Figures 1 and 2), so we repositioned these landmarks using Baidu Maps and Open Street Maps. 

\subsection*{Kernel Density Estimation (KDE)}
\citet{StoyanChiu2024} used the \texttt{bkde2D} function from the \texttt{KernSmooth} package for 2-dimensional kernel smoothing  \citep{KernSmoothPackage}, with bandwidth $h = 3000$~m. \citet{StoyanChiu2024} used a circular Gaussian kernel, i.e. the same bandwidth value in the horizontal and vertical directions. Given that the choice of a bandwidth can be a bit arbitrary, we used the \texttt{bw.diggle} function for automatic bandwidth selection (\texttt{spatstat} package \citep{spatstat}), and obtained a value of $\hall$~m (i.e., yielding less smoothing).  

\citet{Worobey2022Science}, in the peer-reviewed version of their work, used a different KDE function: the \texttt{kde} function from the \texttt{ks} package \citep{ksPackage}. Like \WHSM{} did, we automatically selected a bandwidth matrix using the package's \texttt{Hpi} function. The matrix $H$ affects the amount and the orientation of smoothing. This matrix is not necessarily (and is not, in our case) diagonal, i.e. it is not necessarily aligned along the coordinate axes. 

\subsection*{Bootstrap analyses}

When implementing \citet{StoyanChiu2024}'s test, we estimate $p$~values using resampled datasets of \ecl. We draw the same number of data points as in the original datasets, with replacement, $\nreps$ times. The $p$ value is given by the relative rank of the distance between the original centre and the landmark of interest, compared to the distances between the original centre and the centres of the resampled datasets. 

\bibliographystyle{bibstyle}
\bibliography{biblio_clean}

\clearpage

\appendix
\section*{Appendix}
\renewcommand{\thefigure}{S\arabic{figure}}
\setcounter{figure}{0}

%-------------------------------------------------------------------------------
% Figure population density

\begin{figure}[h!]
\centerfloat
\includegraphics[width = \wpic]{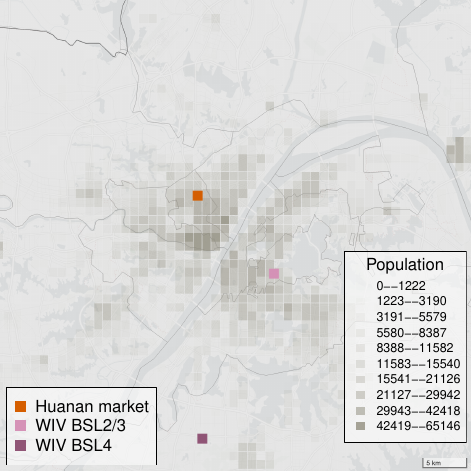}
\caption{Heatmap of population density in Wuhan, manually digitised from \citet{Peng2020Pop}'s Figure~10. }
\label{fig:popdens}
\end{figure}

%-------------------------------------------------------------------------------
% Position of the resampled centres

\begin{figure}
\centerfloat
\begin{tabular}{ll}
$(a)$ With population density data & $(b)$ With Jan Weibo cases \\
\includegraphics[width = \wpic]{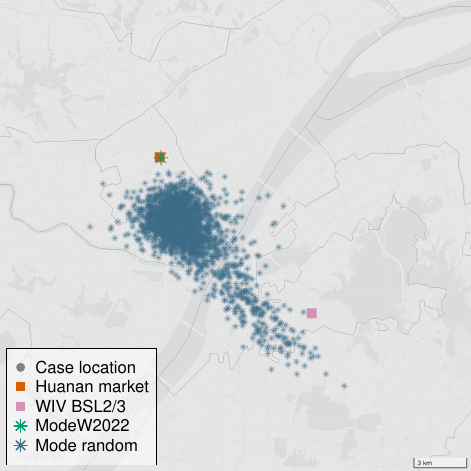} & 
\includegraphics[width = \wpic]{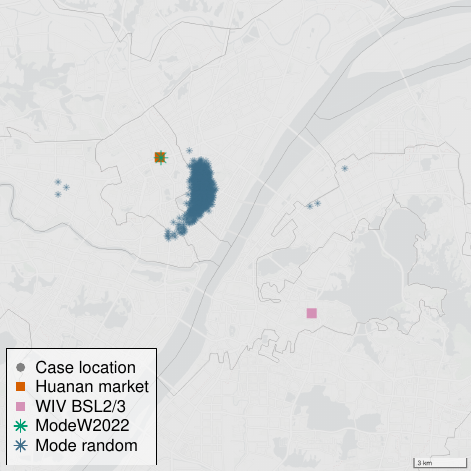}
\end{tabular}
\caption{Positions of the modes of sets of $\ncases$ locations, drawn from $(a)$ Wuhan's population density or $(b)$ from January 2020 Weibo cases; $\nreps$ resampled sets.}
\label{fig:modesRandom}
\end{figure}

%-------------------------------------------------------------------------------
% Rotation centre-point

\begin{figure}[h!]
\centerfloat
\includegraphics[width = \wpic]{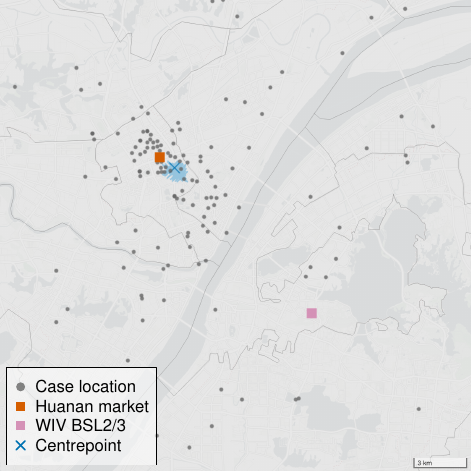}
\caption{Positions of the centre-point of \ecl{} when the coordinates axes are rotated. The centre-point with regular latitude-longitude axes is the centre of rotation, and we tested rotations of $101$ evenly spaced angles between $0$ and $2\pi$; the resulting centre-points are shown in lighter colour. }
\label{fig:rotationcentre-point}
\end{figure}

%-------------------------------------------------------------------------------
% Distribution W2022 cases lat and long

\begin{figure}[h!]
\centerfloat
\begin{tabular}{ll}
$(a)$ Distribution of longitudes of the \ecl{} \\
\includegraphics[]{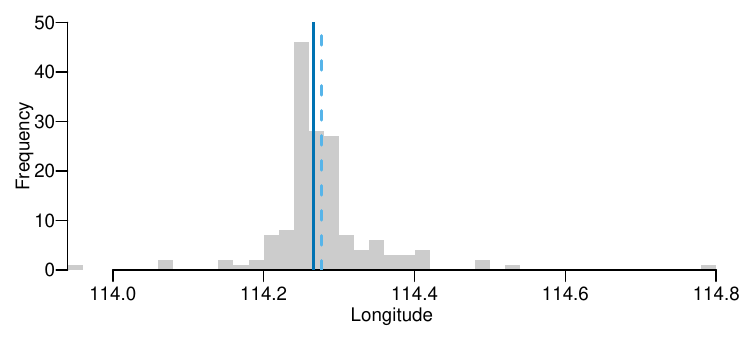}\\
$(b)$ Distribution of latitudes of the \ecl{} \\
\includegraphics[]{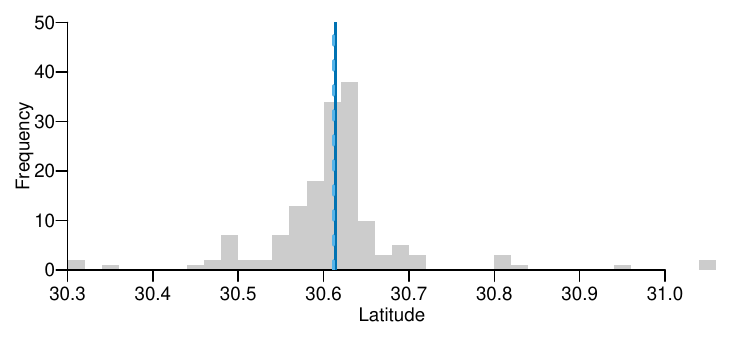}
\end{tabular}
\caption{Distribution of longitudes and latitudes of the \ecl{} in the dataset used by \citet{Worobey2022Science}. The vertical lines correspond to the means (dashed lines) and medians (full lines) of the distributions.}
\label{fig:distributionslonlat}
\end{figure}

%-------------------------------------------------------------------------------
% Zoom out, with Hubei cases

\begin{figure}
\centerfloat
\includegraphics[width = \wpic]{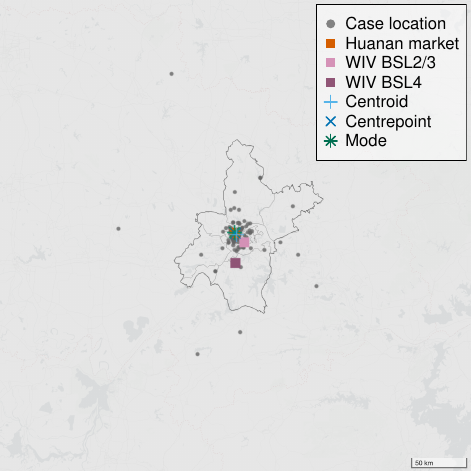}
\caption{All case residential locations, including seven Hubei case locations outside of Wuhan (the city contour is delimited in black).}
\label{fig:NewCasesZoomOut}
\end{figure}

%-------------------------------------------------------------------------------
% Addition of new cases

\begin{figure}[h]
\centerfloat
\begin{tabular}{ll}
$(a)$ Centre-point, cases inside of Wuhan & $(b)$ Centre-point, adding the seven Hubei locations \\[-0.cm] 
\includegraphics[width = \wpic]{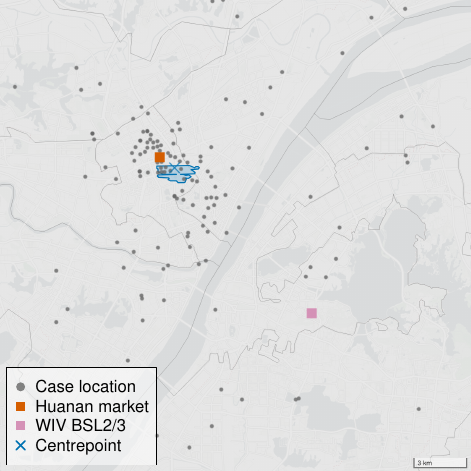}
&
\includegraphics[width = \wpic]{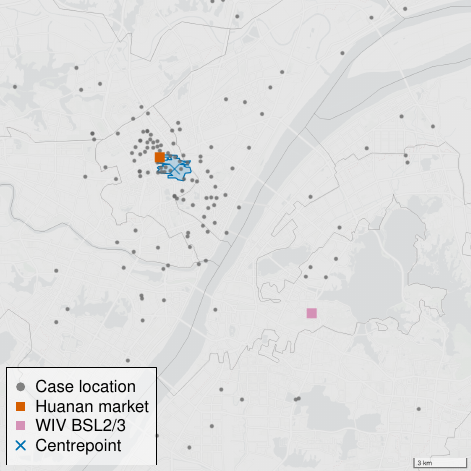}\\
$(c)$ Centroid, cases inside of Wuhan & $(d)$ Centroid, adding the seven Hubei locations \\[-0.cm] 
\includegraphics[width = \wpic]{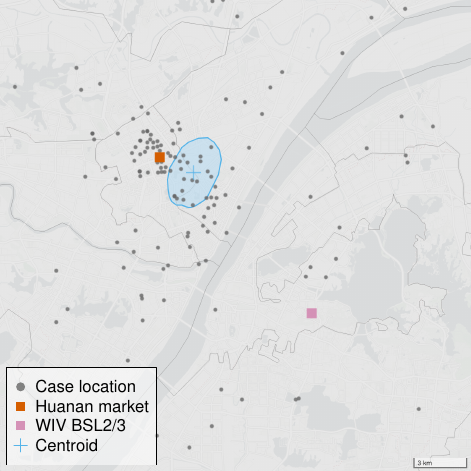}
&
\includegraphics[width = \wpic]{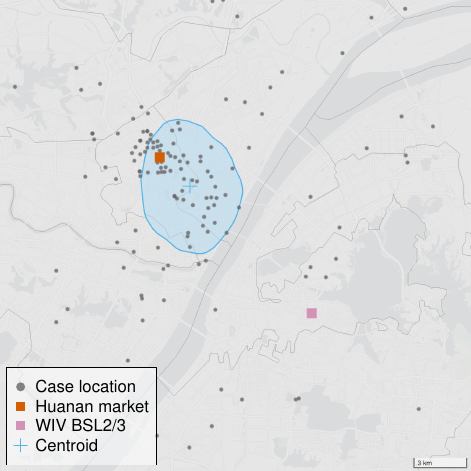}\\
\end{tabular}
\caption{Effect of the addition of new cases on the position of the centre-point and centroid and on their $95\%$ confidence surfaces. The full distribution of cases is shown in Figure~\ref{fig:NewCasesZoomOut}. The contour delimitates the $95\%$ closest bootstrapped centre locations. 
$p$~values: 
$(a)$ centre-point, cases inside Wuhan, $\pvalWcentrepointother$; 
$(b)$ centre-point, with Hubei cases, $\pvalwithOutsidecentrepointoriginal$.
$(c)$ centroid, cases inside Wuhan, $\pvalWcentroidother$; 
$(d)$ centroid, with Hubei cases, $\pvalwithOutsidecentroidoriginal$; $\nreps$ resamples of the data. 
}
\label{fig:WuhanNewCases}
\end{figure}

%-------------------------------------------------------------------------------
% Bandwidth effects, Wuhan, contours
\begin{figure}[h!]
\centerfloat
\begin{tabular}{ll}
$(a)$ \SC's KDE, $h = 3000$~m & $(b)$ \WHSM's KDE, bandwidth matrix $H$ \\
\includegraphics[width = \wpic]{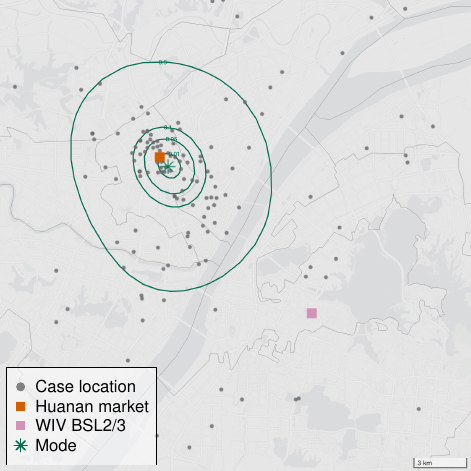}
&
\includegraphics[width = \wpic]{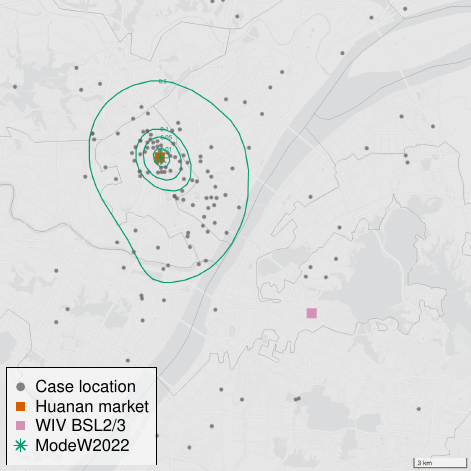} \\
$(c)$ \SC's KDE,  $h = \hall$~m & 
\\
\includegraphics[width = \wpic]{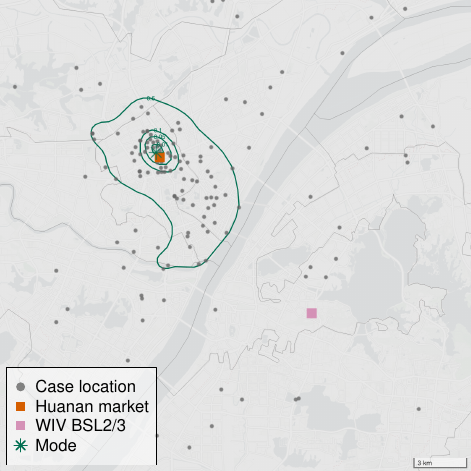} 
&
\end{tabular}
\caption{The impacts of KDE functions and bandwidth choice ($h$) on kernel density estimation
, Wuhan data. \SC{} used $h = 3000$ but a coarser grid, $501 \times 501$, than ours ($1001 \times 1001$), which is why the mode is at the exact same position (but is still very similar; compare the positions of the modes in their Figure 3a to our Figure~\ref{fig:Wuhan_bandwidth}a). In $(b)$, the bandwidth matrix~	$H$ is automatically determined by the \texttt{Hpi} function; in $(c)$ the bandwidth value $h$ is automatically determined by the \texttt{bw.diggle} function. 
}
\label{fig:Wuhan_bandwidth}
\end{figure}

%-------------------------------------------------------------------------------
% Zoom on the market with W2022 mode
\begin{figure}
\centerfloat
\begin{tabular}{ll}
$(a)$ Satellite view & $(b)$ Close-up 
\\
\includegraphics[width = \wpic]{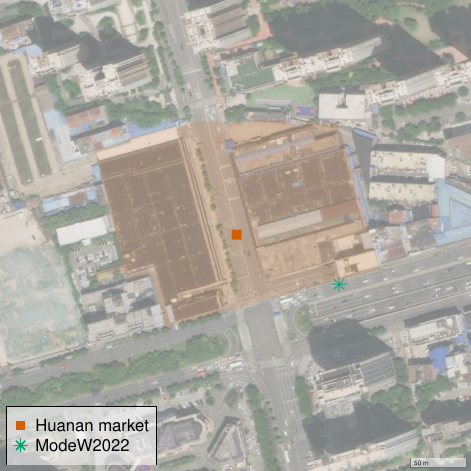} &
\includegraphics[width = \wpic]{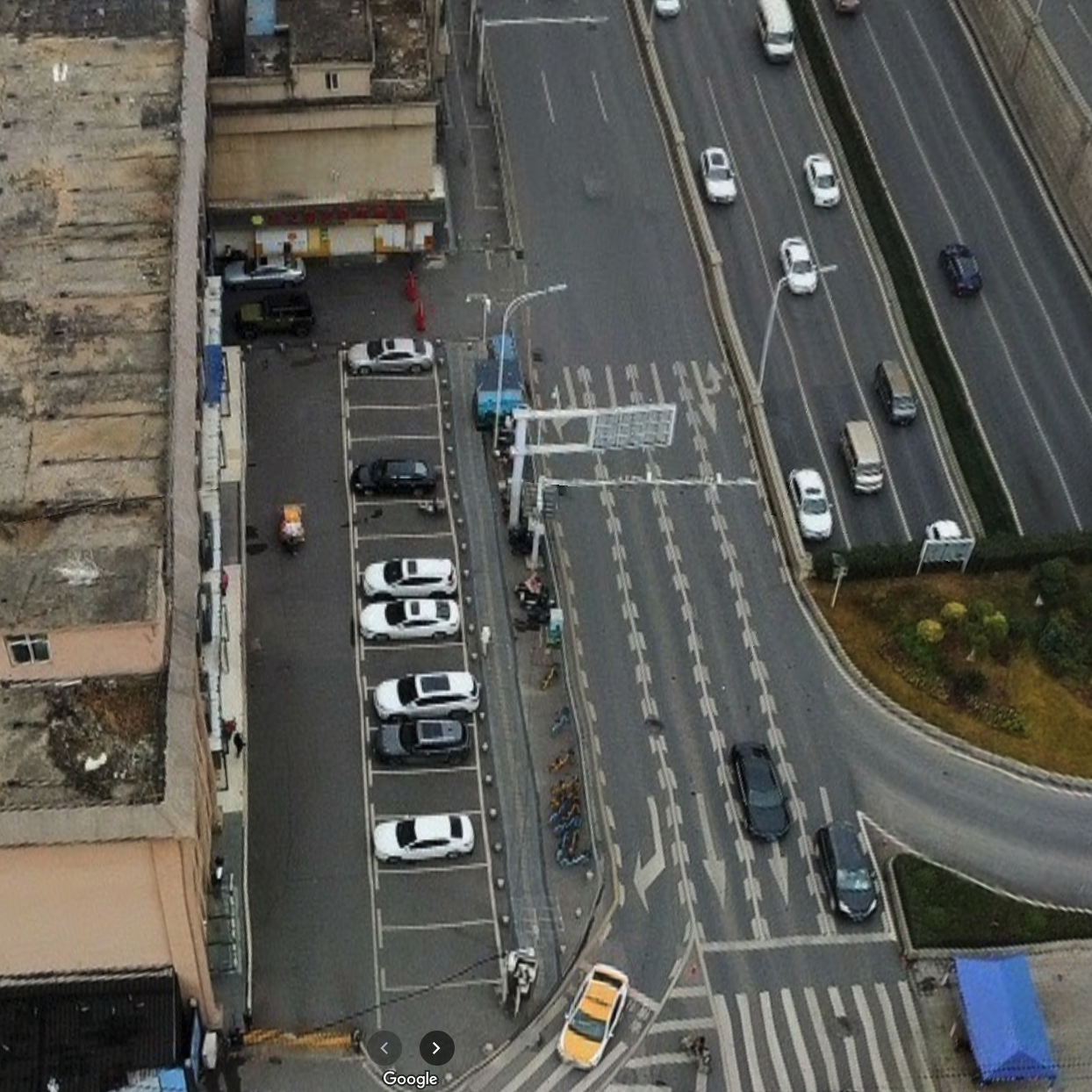}
\end{tabular}
\caption{Zoom on the position of the mode, computed with \WHSM's KDE function, relative to the Huanan market. In panel $(a)$, the red marker between the two sides of the market is the position used throughout our study. Panel $(b)$ further zooms in on the position of the mode (green star in $(a)$); the snapshot is from a picture taken in December 2021, uploaded by user ``logan logan'' on Google Maps, and available as dynamical image at \url{https://maps.app.goo.gl/q5CmZjrTS3bkSGqW8}. }
\label{fig:WuhanZoomMarketMode}
\end{figure}

%-------------------------------------------------------------------------------
% Mode KDE results, SC method 

\begin{figure}[h]
\centerfloat
\begin{tabular}{ll}
$(a)$ \SC's KDE, $h = 3000$ & $(b)$ \SC's KDE, $h = \hall$ \\[-0.cm] 
\includegraphics[width = \wpic]{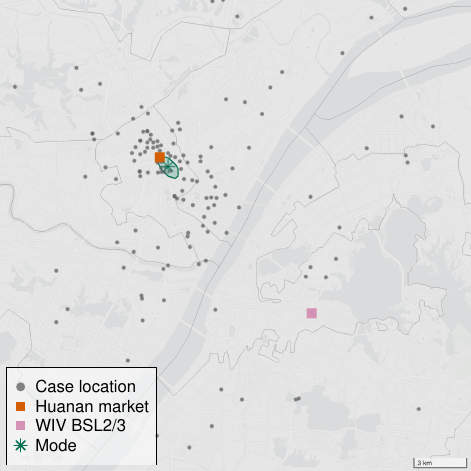}
&
\includegraphics[width = \wpic]{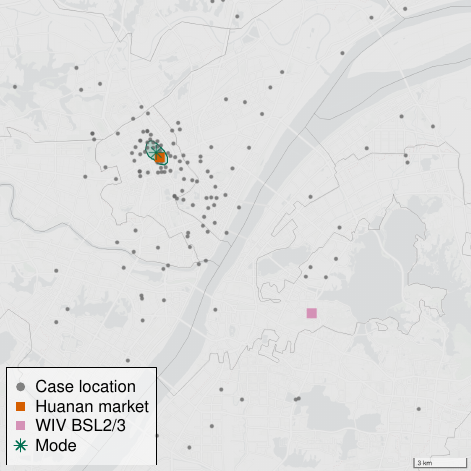} 
\end{tabular}
\caption{Position of the mode and the associated $95\%$ confidence surface, inferred by computing the centres of bootstrap pseudoreplicates, as a function of the bandwidth $h$, using \SC's KDE function implementation in \texttt{R}. Panel $(a)$: $h = 3000$, which is the bandwidth value used by \SC{} ($\pvalWmodeoriginal$). Panel $(b)$: $h = \hall$; the bandwidth value was chosen automatically via the \texttt{bw.diggle} function from the \texttt{spatstat} package \citep{spatstat} ($\pvalWmodeother$). 
The $95\%$ confidence surfaces were determined after $\nreps$ bootstrap resamples in each panel.
We used a finer discretisation grid than \SC{} ($1001\times1001$ instead of $501\times501$).
}
\label{fig:WuhanModeSC}
\end{figure}

%-------------------------------------------------------------------------------
% Landmark locations
\begin{figure}
\centerfloat
\includegraphics[width = \wpic]{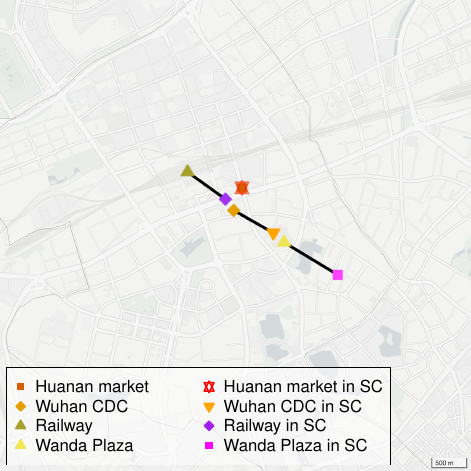}
\caption{Positions of landmarks in Wuhan, using latitude and longitude coordinates shared by \SC{} (``SC'' on the figure) as supplementary information, and showing the corrected positions. As an example, the Hankou railway station is in reality to the North-West of the Huanan market (as shown on \SC's Figure~2 map). It appears in the South-West of the market on their Figure~1 and here, because their coordinates did not take into account the offset on China maps. The position of Wuhan CDC in the figure is different from \SC's Figure 1b, because we used their latitude and longitude coordinates, which are not consistent with the UTM coordinates in \SC's data (UTM coordinates for Wuhan CDC seem to have been manually changed in \SC's data.)}
\label{fig:landmarks}
\end{figure}

%-------------------------------------------------------------------------------
% Landmark locations
\begin{figure}
\centerfloat
\includegraphics[width = \wpic]{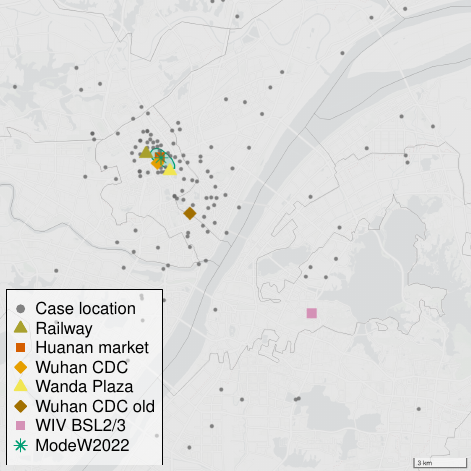}
\caption{Equivalent of Figure~\ref{fig:WuhanMode}a, showing additional landmarks (with locations corrected compared to \SC, see Figure~\ref{fig:landmarks}). }
\label{fig:alllandmarks}
\end{figure}

\end{document}